\documentclass[prl,twocolumn]{revtex4-2}

\usepackage{amsmath}
\usepackage{amssymb}
\usepackage{graphicx}
\usepackage{hyperref}
\usepackage{newtxtext}
\usepackage[varvw]{newtxmath}
\usepackage{bbm}
\usepackage{physics}

\hypersetup{
    colorlinks,
    linkcolor={blue},
    citecolor={blue},
    urlcolor={blue}
}

\setcitestyle{notesep={}}

\graphicspath{{./}{./figs/}}

\newcommand{\rmi}{\mathrm{i}}
\newcommand{\rmd}{\mathrm{d}}

\DeclareMathAlphabet{\mymathbb}{U}{BOONDOX-ds}{m}{n}

\date{February 20, 2024}
\usepackage[dvipsnames]{xcolor}
\usepackage[normalem]{ulem}

\begin{document}

\title{Quasiuniversality from all-in-all-out Weyl quantum criticality in pyrochlore iridates}

\author{David J. Moser}

\author{Lukas Janssen}

\affiliation{Institut f\"ur Theoretische Physik and W\"urzburg-Dresden Cluster of Excellence ct.qmat, TU Dresden, 01062 Dresden, Germany}

%%%%%%%%%%%%%%%%%%%%%%%%%%%%%%%%%%%%%%%%%%%%%%%%%%%%%%%%%%%%%%%%%%%%%%%
\begin{abstract}
We identify an exotic quasiuniversal behavior near the all-in-all-out Weyl quantum critical point in three-dimensional Luttinger semimetals, such as the pyrochlore iridates $R_2$Ir$_2$O$_7$, with $R$ a rare-earth element. The quasiuniversal behavior is characterized by power laws with exponents that vary slowly over several orders of magnitude in energy or length. However, in contrast to the quasiuniversality discussed in the context of deconfined criticality, the present case is characterized by a genuinely-universal ultra-low-temperature behavior. In this limit, the pertinent critical exponents can be computed exactly within a renormalization group analysis. Experimental implications for the pyrochlore iridates are outlined.
\end{abstract}
%%%%%%%%%%%%%%%%%%%%%%%%%%%%%%%%%%%%%%%%%%%%%%%%%%%%%%%%%%%%%%%%%%%%%%%

\maketitle

%%%%%%%%%%%%%%%%%%%%%%%%%%%%%%%%%%%%%%%%%%%%%%%%%%%%%%%%%%%%%%%%%%%%%%%
%\paragraph{Introduction.}
%%%%%%%%%%%%%%%%%%%%%%%%%%%%%%%%%%%%%%%%%%%%%%%%%%%%%%%%%%%%%%%%%%%%%%%
%
Quasiuniversality refers to a situation in which observables display apparent critical behavior over several orders of magnitude of energy or length, where, however, closer inspections of the observed power laws reveal slow drifts in the corresponding exponents~\cite{wang17, nahum20, ma20}.
Such a situation has recently been intensely discussed in the context of deconfined quantum criticality between antiferromagnetic and valence-bond-solid orders in two-dimensional quantum magnets~\cite{senthil04a, senthil04b}. In this case, the quasiuniversal behavior observed in numerical simulations~\cite{nahum15} is believed to arise from a collision and subsequent annihilation of the corresponding critical fixed point with another, bicritical, fixed point, leaving behind a slow renormalization group flow~\cite{ihrig19, wang17, nahum20, ma20}. In this scenario, the ultra-low-temperature behavior is ultimately weakly first order, with, however, a large finite-temperature regime characterized by quasiuniversality.
An alternative interpretation of the numerical data is the presence of a second divergent length scale at criticality~\cite{shao16}. In this competing scenario, the low-temperature behavior is genuinely universal, but requires an adapted scaling ansatz accommodating the presence of the additional length scale.

\begin{figure}[b]
\includegraphics[width=\linewidth]{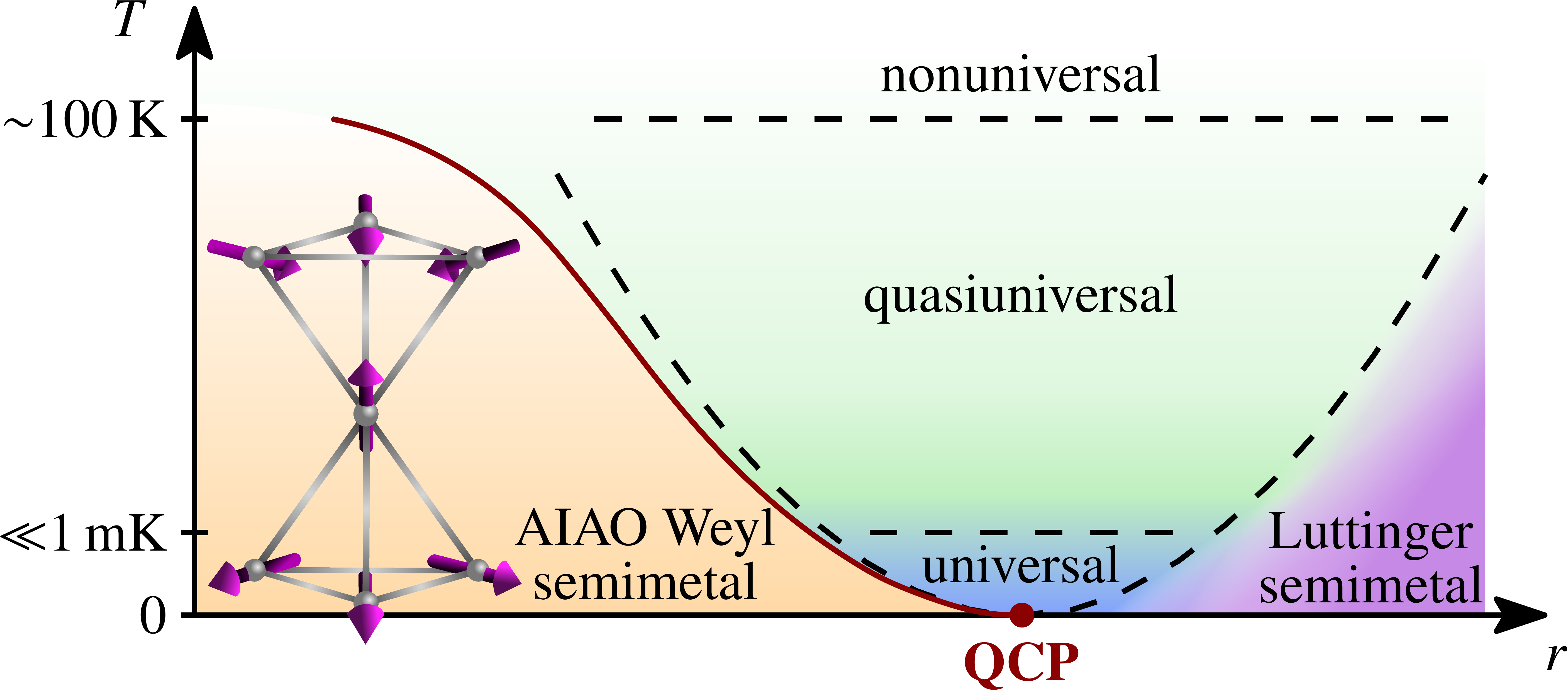}
\caption{Schematic finite-temperature phase diagram of pyrochlore iridates $R_2$Ir$_2$O$_7$ near the quantum critical point (QCP) between the Weyl semimetal with all-in-all-out (AIAO) order, sketched in the inset, and the symmetric Luttinger semimetal.
$r$ indicates a nonthermal tuning parameter, such as chemical doping or hydrostatic pressure.
Between the nonuniversal high-temperature regime and the genuinely-universal ultra-low-temperature regime, there is a large quasiuniversal regime, characterized by power laws with slowly varying exponents.}
\label{fig:intro}
\end{figure}

In the present work, we identify quasiuniversal behavior in a three-dimensional model relevant to a class of pyrochlore iridates  with chemical composition $R_2$Ir$_2$O$_7$, where $R$ is a rare-earth element, e.g., $R = \mathrm{Pr}, \mathrm{Nd}$.
In their metallic phases, these compounds exhibit an electronic excitation spectrum characterized by quadratic band touching at the Fermi level~\cite{kondo15, nakayama16}, and as such fall into the larger class of Luttinger semimetals~\cite{armitage18, lv21}.
However, in comparison with the prominent members of this class, HgTe~\cite{bruene11} and $\alpha$-Sn~\cite{barfuss13}, the pyrochlore iridates feature a substantially increased effective quasiparticle mass, implying an enhanced role of electronic interactions~\cite{cheng17, wang20}.
If strong enough, these interactions can drive symmetry-breaking transitions, across which the electronic spectrum becomes partially or fully gapped out.
Many pyrochlore iridates, such as Nd$_2$Ir$_2$O$_7$~\cite{ma15, nakayama16}, indeed display a finite-temperature transition, below which the iridium moments feature all-in-all-out (AIAO) antiferromagnetic order. This state breaks time reversal but preserves crystal symmetries~\cite{witczak14}.
Pr$_2$Ir$_2$O$_7$, by contrast, appears to remain disordered up to the lowest accessible temperatures~\cite{nakatsuji06, cheng17}.
By varying the concentration~$x$ in (Pr$_{x}$Nd$_{1-x}$)$_2$Ir$_2$O$_7$ and/or by applying hydrostatic pressure, the N\'eel temperature associated with the onset of AIAO order can be tuned to zero, uncovering an underlying quantum phase transition~\cite{ueda17}.
Small AIAO order converts the quadratic band touching point into eight symmetry-related linear band crossing points~\cite{witczak12a, witczak12b, berke18}.
Consequently, the quantum phase transition is expected to separate the symmetric Luttinger semimetal from a time-reversal-broken Weyl semimetal.
The presence of gapless fermions at the transition indicates the possibility of unconventional behavior.
In fact, previous theoretical work suggested a novel type of fermionic quantum critical point~\cite{savary14, boettcher17}.
Here, we elucidate the finite-temperature properties of this transition, relevant for the experiments on the pyrochlore iridates.  We reveal a large finite-temperature regime above the quantum critical point that is characterized by quasiuniversality, see Fig.~\ref{fig:intro}.
%
%%%%%%%%%%%%%%%%%%%%%%%%%%%%%%%%%%%%%%%%%%%%%%%%%%%%%%%%%%%%%%%%%%%%%%%
%%%%%%%%%%%%%%%%%%%%%%%%%%%%%%%%%%%%%%%%%%%%%%%%%%%%%%%%%%%%%%%%%%%%%%%
%
This unusual behavior arises from the presence of a marginally-irrelevant coupling at the corresponding renormalization group fixed point.
Importantly, we present an approach that allows us to identify properties of this fermionic quantum critical point exactly, including its nontrivial critical exponents.
Our results reveal the AIAO Weyl quantum critical point in pyrochlore iridates as a unique instance of an interacting continuous quantum phase transition that is both experimentally and theoretically accessible.
As a side product, we demonstrate the emergence of strong cubic anisotropy at criticality, thereby resolving an apparent contradiction in the literature~\cite{savary14, boettcher17}.

%%%%%%%%%%%%%%%%%%%%%%%%%%%%%%%%%%%%%%%%%%%%%%%%%%%%%%%%%%%%%%%%%%%%%%%
\paragraph{Model.}
%%%%%%%%%%%%%%%%%%%%%%%%%%%%%%%%%%%%%%%%%%%%%%%%%%%%%%%%%%%%%%%%%%%%%%%
%
At the quantum critical point, the system can be effectively described by a continuum Euclidean action $S = \int \dd{\tau} \dd[3]{x} L$, with Lagrangian $L = L_0 + L_a + L_\phi$. 
Here, $L_0$ describes noninteracting electronic quasiparticles near the quadratic band touching~\cite{abrikosov71, abrikosov74, moon13, herbut14, janssen15, murakami04},
\begin{align} \label{EqLagrangianFree}
    L_0 = \sum_{i=1}^N \psi^{\dagger}_i \left( \partial_\tau + \sum_{a = 1}^5 (1 + s_a \delta) d_a(-\rmi \nabla) \gamma_a \right) \psi_i, 
\end{align}
and originates in the Luttinger Hamiltonian~\cite{luttinger56}.
In the above,
$N$ corresponds to the number of band touching points at the Fermi level, with $N=1$ in the case of the pyrochlore iridates,
$\psi_i$ is a four-component Grassmann field, 
$s_a \coloneqq 1$ ($s_a \coloneqq -1$) for $a = 1, 2, 3$ ($a = 4,5$),
$\delta$ parametrizes the cubic anisotropy with $-1 \leq \delta \leq 1$,
the $4 \times 4$ matrices $\gamma_a$ fulfill the Euclidean Clifford algebra, $\{ \gamma_a ,\gamma_b \} = 2 \delta_{ab} \mathbbm 1$,
and $d_a$ are proportional to $\ell = 2$ real spherical harmonics, viz.\ $d_1 (\vec{p}) = \sqrt{3} p_y p_z$, $d_2 (\vec{p}) = \sqrt{3} p_x p_z$, $d_3 (\vec{p}) = \sqrt{3} p_x p_y$, $d_4 (\vec{p}) = \frac{\sqrt{3}}{2} (p_x^2 - p_y^2)$, and $d_5 (\vec{p}) = \frac{1}{2} (2 p_z^2 - p_x^2 - p_y^2)$.
Moreover, we account for the long-range Coulomb interaction $\sim 1/|\vec x|$ via~\cite{herbut14, janssen16, janssen17}
\begin{align} \label{EqLagrangianPhoton}
    L_a = \frac{1}{2} (\nabla a)^2 + \rmi e a \sum_{i=1}^N \psi^{\dagger}_i \psi_i,
\end{align}
where $a$ denotes the scalar Coulomb field, and $e$ the effective charge. 
Fluctuations corresponding to AIAO ordering on the pyrochlore lattice are parametrized by an Ising field $\phi$, and couple to the electronic quasiparticles as~\cite{savary14, boettcher17, ladovrechis21}
\begin{align} \label{EqLagrangianOP}
    L_\phi = \frac{1}{2} \phi ( r -\nabla^2 ) \phi + g \phi \sum_{i=1}^N \psi^{\dagger}_i \gamma_{45} \psi_i .
\end{align}
Here, $r$ denotes the tuning parameter for the quantum phase transition, $g$ the coupling constant, and $\gamma_{45} \coloneqq \rmi \gamma_4 \gamma_5$.
A term $(\partial_\tau \phi)^2$ can be included in Eq.~\eqref{EqLagrangianOP} as well, but is power-counting irrelevant and as such does not change the critical behavior. The same is true for bosonic self-interactions, such as a $\phi^4$ term.
A finite expectation value $\langle \phi \rangle \neq 0$ breaks time reversal and splits the quadratic band touching point into four pairs of Weyl nodes along the $[111]$ and symmetry-related axes in the cubic basis.

\begin{figure}[tb]
\includegraphics[width=\linewidth]{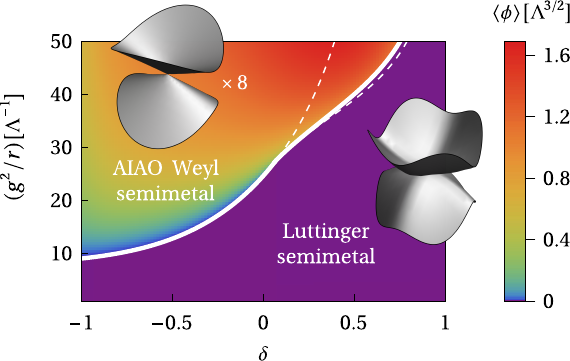}
\caption{Zero-temperature phase diagram of effective model as function of cubic anisotropy $\delta$ and interaction strength $g^2/r$, from mean-field theory.
Coloring indicates magnitude of AIAO order parameter~$\langle \phi \rangle$. 
With the onset of order, the quadratic band touching point of the Luttinger semimetal splits into four pairs of Weyl points in the AIAO Weyl semimetal, see insets.
The transition is continuous (discontinuous) for $\delta \leq \delta_0$ ($\delta > \delta_0$), with $\delta_0 \approx 0.0624$.
Dashed lines delimit the region in which metastable states exist.}
\label{fig:MFT}
\end{figure}

%%%%%%%%%%%%%%%%%%%%%%%%%%%%%%%%%%%%%%%%%%%%%%%%%%%%%%%%%%%%%%%%%%%%%%%
\paragraph{Mean-field analysis.}
%%%%%%%%%%%%%%%%%%%%%%%%%%%%%%%%%%%%%%%%%%%%%%%%%%%%%%%%%%%%%%%%%%%%%%%
%
We start by discussing the model on the level of mean-field theory at zero temperature.
Formally, mean-field theory corresponds to the limit of large number $N$ of quadratic band touching points at the Fermi level. This effectively suppresses fluctuations of the bosonic fields.
The value of the order parameter is then obtained by minimizing the mean-field energy
$E_\text{MF} (\phi) = \frac{r}{2} \phi^2 + \sum_{i=1}^{2} \int_{\vec p} \varepsilon_{\phi}^{(i)}(\vec p)$,
where $\varepsilon_{\phi}^{(1,2)}$ denote the two lower-branch eigenvalues of the mean-field Hamiltonian
$H_\text{MF} = \sum_{a = 1}^5 (1 + s_a \delta) d_a(\vec{p}) \gamma_a + g \phi \gamma_{45}$,
see Supplemental Material (SM) for details~\cite[(a)]{supplemental}.
The resulting phase diagram is presented in Fig.~\ref{fig:MFT}.
%
%%%%%%%%%%%%%%%%%%%%%%%%%%%%%%%%%%%%%%%%%%%%%%%%%%%%%%%%%%%%%%%%%%%%%%%
%%%%%%%%%%%%%%%%%%%%%%%%%%%%%%%%%%%%%%%%%%%%%%%%%%%%%%%%%%%%%%%%%%%%%%%
%
We observe two distinct phases, the paramagnetic Luttinger semimetal phase and the time-reversal-broken Weyl semimetal phase. The former is located at small coupling below a finite threshold $(g^2/r)_\text{c}$ and hosts a quadratic band touching point with four-fold degeneracy at zero momentum.
The Weyl semimetal phase, characterized by a finite order parameter, is encountered above the phase boundary.
It hosts four pairs of Weyl nodes along the $[111]$ and symmetry-related axes in the electronic spectrum, and features AIAO order on the pyrochlore lattice.
For anisotropies $\delta \leq \delta_0 \approx 0.0624$, we observe a continuous phase transition, while for $\delta > \delta_0$, the phase transition becomes discontinuous. The latter case gives rise to metastable states in the region around the transition, delimited by the dashed lines in Fig.~\ref{fig:MFT}.
Importantly, ab-initio calculations and photo emission spectroscopy experiments in Pr$_2$Ir$_2$O$_7$~\cite{kondo15} suggest $\delta<0$, placing this material into the regime with continuous transition~\cite{note_delta}.
Note that at strong coupling, a second transition towards a Mott-insulating phase may be expected~\cite{wan11, witczak12a, berke18}, which is not captured by the mean-field theory of our effective model.
However, the presence or not of this strong-coupling phase is irrelevant for the physics close to the quantum critical point between Luttinger and AIAO Weyl semimetals, which we will focus on in the following.

%%%%%%%%%%%%%%%%%%%%%%%%%%%%%%%%%%%%%%%%%%%%%%%%%%%%%%%%%%%%%%%%%%%%%%%
\paragraph{Renormalization group analysis.}
%%%%%%%%%%%%%%%%%%%%%%%%%%%%%%%%%%%%%%%%%%%%%%%%%%%%%%%%%%%%%%%%%%%%%%%

%%%%%%%%%%%%%%%%%%%%%%%%%%%%%%%%%%%%%%%%%%%%%%%%%%%%%%%%%%%%%%%%%%%%%%%
\begin{figure*}[tb]
\includegraphics[width=\linewidth]{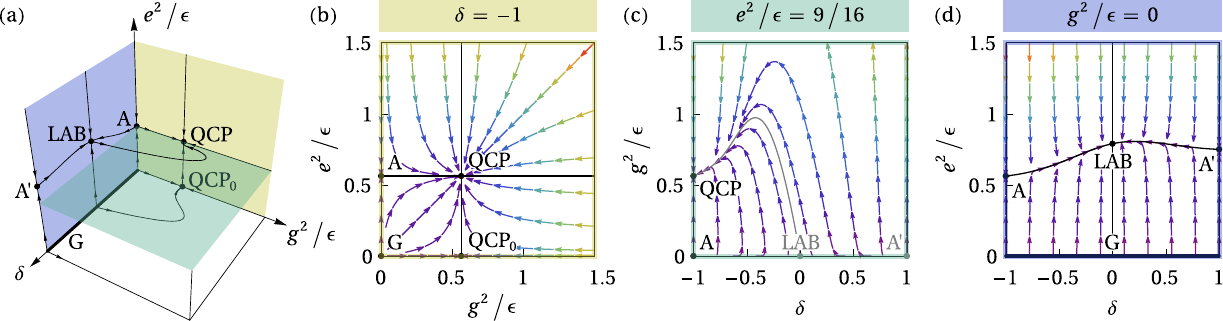}
\caption{%
Fixed-point structure for $N=1$ on the critical hypersurface $r=0$ in 
(a) the parameter space spanned by $\delta$, $g^2$, and $e^2$, 
(b) the $g^2$-$e^2$ plane for fixed $\delta = -1$, 
(c) the $\delta$-$g^2$ plane for fixed $e^2 = 9\epsilon/16$, and 
(d) the $\delta$-$e^2$ plane for fixed $g^2 = 0$.
Black (gray) dots and lines indicate fixed points and separatrices, respectively, located within (projected onto) the respective planes.
Arrows denote flow towards the infrared, with their coloring indicating the flow velocity.
There is a unique quantum critical fixed point at finite coupling, labeled as QCP. It corresponds to the continuous transition between the symmetric Luttinger semimetal and the AIAO-ordered Weyl semimetal.
The symmetric state is described by the Luttinger-Abrikosov-Beneslavskii fixed point at $g^2 = 0$, labeled as LAB.
The dots labeled as A, A', QCP$_0$, and G correspond to unstable interacting and Gaussian, respectively, fixed points, further discussed in the SM.}
\label{fig:flow-diagram}
\end{figure*}
%%%%%%%%%%%%%%%%%%%%%%%%%%%%%%%%%%%%%%%%%%%%%%%%%%%%%%%%%%%%%%%%%%%%%%%

Note that the partially bosonized Lagrangian $L = L_0 + L_a + L_\phi$ features a unique upper critical dimension, since both the Yukawa coupling $g$ and the effective charge $e$ become marginal in $d = 4$ spatial dimensions. This allows a standard $\epsilon = 4 - d$ expansion.
We start by discussing the results at one-loop order, naively valid only for small $\epsilon$, but will then show that higher-loop corrections in fact vanish exactly at the AIAO Weyl quantum critical point.
Integrating out modes with momenta $q$ in the thin shell $\Lambda/b < q < \Lambda$, where $\Lambda$ denotes the ultraviolet cutoff, and arbitrary frequency, leads to the flow equations at criticality $r=0$ as~\cite[(b)]{supplemental}
\begin{align}
    \frac{\rmd\delta}{\rmd \ln b} & = 
    - \frac{2}{15}(1-\delta^2)^2
    \left[
    (g^2 + e^2) f_{1 \text{e}} + (g^2 - e^2) f_{1 \text{t}}
    \right],
\label{eq:betaFunctionDelta} \allowdisplaybreaks[1] \\
    \frac{\rmd g^2}{\rmd \ln b} & = (\epsilon - \eta_\phi)g^2 
    - \frac{2}{15}(1-\delta^2)\bigl[
    (1+\delta) (g^2+e^2) f_{1\text{e}}
    \nonumber\\&\quad
    - (1-\delta)(g^2-e^2)(f_{1\text{t}}-3f_{2\text{e}})
    \bigr]g^2, 
\label{eq:betaFunctionG2} \allowdisplaybreaks[1] \\
    \frac{\rmd e^2}{\rmd \ln b} & = (\epsilon - \eta_a)e^2
    - \frac{2}{15}(1-\delta^2)\bigl[
    (1+\delta) (g^2+e^2) f_{1\text{e}}
    \nonumber\\&\quad
    - (1-\delta) (g^2-e^2) f_{1\text{t}}
    \bigr]e^2,
\label{eq:betaFunctionE2} 
\end{align}
where $\eta_\phi = N g^2 f_{g^2}$ and $\eta_a = N e^2 f_{e^2}$. 
Here, $f_i \equiv f_i(\delta)$ with $i \in \{1\text{e}, 1\text{t}, 2\text{e}, e^2\}$ ($i \in \{g^2\}$) are bounded and continuous functions of the anisotropy parameter $\delta$ with $f_i > 0$ ($f_{g^2} \geq -2/3$) for $\delta \in [-1,1]$ and $f_i = 1$ ($f_{g^2} = 0$) for $\delta = 0$.
Their definitions and numerical values for $\delta \neq 0$ are given in the SM~\cite[(c)]{supplemental}.
In the above flow equations, we have rescaled the couplings as $(g^2,e^2) \Lambda^{-\epsilon}/ (2 \pi^2) \mapsto (1 - \delta^2) (g^2,e^2)$, which turns out convenient to assess the properties of the stable fixed point.
Note that this implies that higher-order loop corrections to the above equations will involve additional factors of $(1-\delta^2)$.
Importantly, in the present form, the stable fixed point associated with AIAO Weyl quantum criticality is located at finite couplings, as we show next.

Figure~\ref{fig:flow-diagram}(a) depicts the fixed-point structure for $N=1$ on the critical hypersurface $r=0$ in the space spanned by the parameters $\delta$, $g^2$, and $e^2$.
Notably, both $g^2$ and $e^2$ are relevant couplings and flow towards a finite value in the infrared, Fig.~\ref{fig:flow-diagram}(b).
For finite $g^2$, the anisotropy parameter $\delta$ does then no longer have a fixed point at $\delta = 0$, and instead flows towards maximal anisotropy $\delta = -1$, Fig.~\ref{fig:flow-diagram}(c).
Most importantly, there is a unique stable fixed point at finite couplings, located at $\delta_\star = -1$ and $g_\star^2 = e_\star^2 = 9\epsilon/(16 N)$.
This fixed point, labeled as QCP in Figs.~\ref{fig:flow-diagram}(a)--\ref{fig:flow-diagram}(c), corresponds to the continuous quantum phase transition between the symmetric Luttinger semimetal, described by the Luttinger-Abrikosov-Beneslavskii fixed point~\cite{abrikosov71, abrikosov74, moon13, herbut14} in the $g=0$ plane, Fig.~\ref{fig:flow-diagram}(d), and the AIAO-ordered Weyl semimetal.
It is characterized by finite boson anomalous dimensions $\eta_\phi = \eta_a = \epsilon$, vanishing fermion anomalous dimension $\eta_\psi = 0$, and a dynamical critical exponent $z=2$.
From the flow of the tuning parameter $r$, we furthermore obtain the correlation-length exponent $\nu = 1/(2-\epsilon)$.
In the SM, we show that particle-hole symmetry becomes emergent at the quantum critical point~\cite[(d)]{supplemental}.
The SM also contains a discussion of the other, unstable, fixed points~\cite[(e)]{supplemental}.

%%%%%%%%%%%%%%%%%%%%%%%%%%%%%%%%%%%%%%%%%%%%%%%%%%%%%%%%%%%%%%%%%%%%%%%
\begin{figure*}
\includegraphics[width=\linewidth]{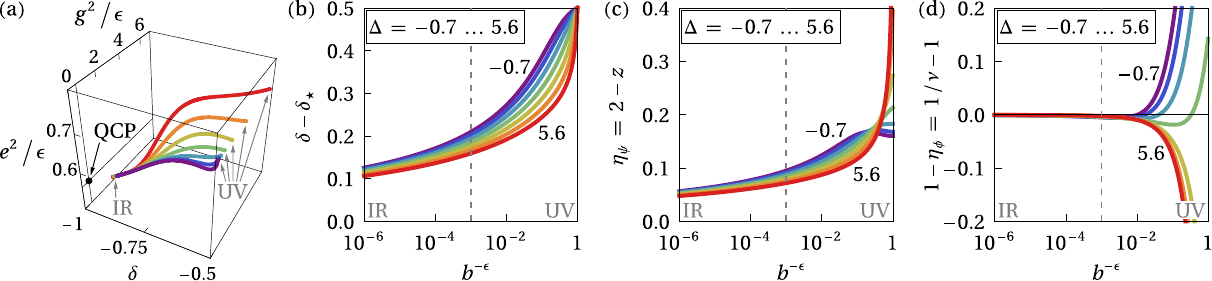}
\caption{%
Renormalization group flow on the critical hypersurface $r=0$ from the ultraviolet scale, corresponding to $b^{-\epsilon} = 1$, to a deep infrared scale, corresponding to $b^{-\epsilon} = 10^{-6}$, starting from different microscopic parameters. 
Each curve corresponds to a numerical integration of the flow equations using the initial conditions $(\delta, g^2, e^2) = (-0.5, 0.7391, 0.9098 + \Delta)$ for $b^{-\epsilon}=1$, with $\Delta = -0.7, -0.6, -0.4, 0, 0.8, 2.4, 5.6$ from purple to red. 
The starting values of $(g^2, e^2)$ have been chosen to satisfy the pseudo-fixed-point conditions for $\delta = -0.5$ when $\Delta = 0$.
(a) Flow trajectories in the parameter space spanned by $\delta$, $g^2$, and $e^2$, illustrating the crossover from the nonuniversal regime, characterized by independent trajectories, to the quasiuniversal regime, in which these collapse onto a single curve.
Deviations of (b) anisotropy parameter $\delta$ and effective exponents (c) $\eta_\psi = 2 - z$ and (d) $\eta_\phi = 2- 1/\nu$ from their respective critical values as function of renormalization group scale $b^{-\epsilon}$.
%
%, with $b^{-\epsilon}=1$ ($b^{-\epsilon} \ll 1$) corresponding to the ultraviolet (infrared) scale.
%
The quasiuniversal regime emerges at $b^{-\epsilon} \lesssim 10^{-3}$, as indicated by the dashed line, and manifests itself in $(g^2, e^2)$-independent, but anisotropy-dependent, drifting exponents.
Note that the flow is still significantly away from its ultra-low-energy limit even at $b^{-\epsilon} = 10^{-6}$, as evidenced by the finite deviation $\delta - \delta_\star \gtrsim 0.1$ in (b).}
\label{fig:quasiuniversality}
\end{figure*}
%%%%%%%%%%%%%%%%%%%%%%%%%%%%%%%%%%%%%%%%%%%%%%%%%%%%%%%%%%%%%%%%%%%%%%%

The stable fixed point has remarkable properties.
First of all, the fact that it is located at $\delta_\star = -1$ implies that higher-loop corrections to the vertex renormalizations, Eqs.~\eqref{eq:betaFunctionG2} and \eqref{eq:betaFunctionE2}, vanish at criticality, see SM for details~\cite[(f)]{supplemental}. As a consequence, we find that our one-loop results $z=2$, $\eta_\phi = \eta_a = 4-d$, and $\nu = 1/(d-2)$, hold at all loop orders at the quantum critical point.
The transition between Luttinger and AIAO Weyl semimetals in the pyrochlore iridates therefore realizes a rare instance of an interacting fermionic quantum critical point in a three-dimensional system that allows an exact determination of its critical properties.
With the exact results at hand, we can resolve an apparent contradiction in the literature concerning the relevance or not of the cubic anisotropy at this quantum critical point~\cite{savary14, boettcher17}.
While the anisotropy parameter $\delta$ vanishes at the Luttinger-Abrikosov-Beneslavskii fixed point, it flows towards $\delta = -1$ at the quantum critical fixed point.
Technically, the emergence of this maximal anisotropy at criticality arises from the order-parameter contributions to the fermion self-energy, which are neglected in the analysis of Ref.~\cite{boettcher17}.

%%%%%%%%%%%%%%%%%%%%%%%%%%%%%%%%%%%%%%%%%%%%%%%%%%%%%%%%%%%%%%%%%%%%%%%
\paragraph{Quasiuniversality.}
%%%%%%%%%%%%%%%%%%%%%%%%%%%%%%%%%%%%%%%%%%%%%%%%%%%%%%%%%%%%%%%%%%%%%%%
%
In the vicinity of the quantum critical fixed point at $\delta_\star = -1$, the flow of the anisotropy parameter takes the form
\begin{align}
    \dv{(\delta-\delta_\star)}{\ln b} \Bigg\arrowvert_{g_\star^2, e_\star^2} = - \frac{c \epsilon}{N} (\delta - \delta_\star)^2 + \mathcal{O} ((\delta - \delta_\star)^3)
\end{align}
with a constant $c \coloneqq 3 f_{1 \text{e}}(-1) / 5 \approx 0.4449$. 
Note the absence of a linear term $\propto (\delta - \delta_\star)$ in the above equation.
The deviation $(\delta - \delta_\star)$ corresponds to a \emph{marginally} irrelevant parameter, implying a logarithmically slow flow towards the critical point,
$\delta (b) - \delta_\star \simeq N/(c \epsilon \ln b)$,
for $\ln b \gg 1$.
By contrast, the Yukawa coupling and the effective charge acquire a power-law flow, $g^2(b) - g^2_\star \propto b^{-\epsilon}$ and $e^2(b) - e^2_\star \propto b^{-\epsilon}$.
This implies a separation of scales, giving rise to the following three regimes in energy:
In the nonuniversal high-energy regime, the couplings $g^2$ and $e^2$ flow from their microscopic values towards pseudo-fixed-point values $g_*^2 (\delta)$ and $e_*^2 (\delta)$, which depend only on the anisotropy parameter $\delta$. Within this regime, the latter can be considered approximately constant.
Its slow flow becomes visible only when several orders of magnitude in energy are considered.
This defines a quasiuniversal intermediate-to-low-energy regime~\cite{nahum15, nahum20}, in which the couplings $g^2$ and $e^2$ do no longer depend on their microscopic values, but solely follow their pseudo-fixed-point values $g_*^2 (\delta)$ and $e_*^2 (\delta)$ with slowly varying anisotropy parameter $\delta$.
In this regime, the correlation length becomes large, such that observables display approximate power laws with slowly drifting exponents~\cite{schwab22, dresselhaus21}.
Finally, the genuinely-universal ultra-low-energy regime will be reached only after the anisotropy parameter has approached its ultimate infrared regime, which requires fluctuations on unusually many energy scales to be integrated out.

This is exemplified in Fig.~\ref{fig:quasiuniversality}, which shows the numerically-integrated renormalization group flow for different initial couplings on the critical hypersurface $r=0$.
The couplings $g^2$ and $e^2$ first exhibit a fast flow on individual, nonuniversal, trajectories, but then approach a single, quasiuniversal, trajectory of pseudo-fixed points $(g_*^2(\delta), e^2_*(\delta))$, along which they only slowly flow, as a consequence of the slow flow of $\delta$, Fig.~\ref{fig:quasiuniversality}(a).
In fact, even after six orders of magnitude in $b^\epsilon$ have been integrated out, $\delta$ is still significantly away from its ultra-low-energy value $\delta_\star = -1$, Fig.~\ref{fig:quasiuniversality}(b).
There is therefore a large quasiuniversal regime characterized by approximate power laws with effective critical exponents, which are only slowly drifting with energy or length. 
As shown in the SM~\cite[(b)]{supplemental}, the present model satisfies additional scaling relations between the different exponent, viz.\ $z = 2 - \eta_\psi$ and $1/\nu  = 2- \eta_\phi$. There are therefore only two independent effective exponents. Their slow drifts as function of scale are depicted in Figs.~\ref{fig:quasiuniversality}(c) and (d), illustrating the fact that the genuinely-universal values are reached, at least for some of the exponents, only in the ultra-low-energy limit.
We emphasize that the slow renormalization group flow is independent of microscopic parameters and as such will occur in any given material realization of the all-in-all-out Weyl quantum critical point in Luttinger semimetals.
In the SM~\cite[(g)]{supplemental}, we compare the quasiuniversal flow with a generic power-law flow of a toy model that does not feature a marginal coupling, illustrating the unusual behavior observed in the quasiuniversal regime.

%%%%%%%%%%%%%%%%%%%%%%%%%%%%%%%%%%%%%%%%%%%%%%%%%%%%%%%%%%%%%%%%%%%%%%%
\paragraph{Experiments.}
%%%%%%%%%%%%%%%%%%%%%%%%%%%%%%%%%%%%%%%%%%%%%%%%%%%%%%%%%%%%%%%%%%%%%%%
%
The quasiuniversal regime is separated by crossovers from the nonuniversal regime at the ultraviolet scale and the genuinely-universal regime in the deep infrared, and realizes a novel strongly-interacting quantum state of matter.
For the pyrochlore iridates, the relevant energy scale above which the physics depends on microscopic details of the material is around $100$~K~\cite{witczak14}. 
Assuming $z \simeq 2$, each order of magnitude in energy corresponds to only half an order of magnitude in $b^\epsilon$, if $\epsilon = 1$.
The ultra-low-temperature behavior will therefore be reached only well below the 1~mK regime. Such regime not only is difficult to access experimentally and necessitates sufficiently pure crystals that do not cut off the required long-range fluctuations: In the case of the pyrochlore iridates, we also expect new effects in this ultra-low-energy regime, arising from the weak Kondo coupling of the iridium electrons to the rare-earth local moments~\cite{yao18, ueda22}.
The experimentally most easily accessible regime below 100~K, by contrast, will be governed by quasiuniversal behavior, characterized by approximate power laws with slowly drifting exponents.
The specific heat as function of temperature, for instance, is expected to scale as $C \sim T^{d/z}$ with drifting exponent $d/z > 3/2$ at around $100$\,K and $d/z = 3/2$ in the ultra-low-temperature limit.
Observing such drifting exponents, e.g., in careful thermodynamic and/or transport measurements, would accomplish the first experimental realization of this new state of matter.

%%%%%%%%%%%%%%%%%%%%%%%%%%%%%%%%%%%%%%%%%%%%%%%%%%%%%%%%%%%%%%%%%%%%%%%
\paragraph{Conclusions.}
%%%%%%%%%%%%%%%%%%%%%%%%%%%%%%%%%%%%%%%%%%%%%%%%%%%%%%%%%%%%%%%%%%%%%%%
%
We have demonstrated the emergence of a novel quasiuniversal regime, which should be understood as a new strongly-interacting quantum state of matter, in the finite-temperature phase diagram of interacting three-dimensional Luttinger semimetals.
While quasiuniversal behavior is usually associated with a fixed-point annihilation scenario~\cite{nahum20, ma20}, our results show that the presence of a marginally-irrelevant operator can lead to the same phenomenology.
Our findings call for new experiments on sufficiently pure samples of pyrochlore iridates $R_2$Ir$_2$O$_7$, which search for power laws with slowly drifting exponents in the intermediate-energy regime above the Weyl-Luttinger quantum phase transition~\cite{ueda17, nikolic22}.
In the ultra-low-energy regime, by contrast, the weak Kondo coupling between the iridium electrons and the rare-earth local moments might lead to even more intriguing effects, the study of which represents an excellent direction for future theoretical work.

%%%%%%%%%%%%%%%%%%%%%%%%%%%%%%%%%%%%%%%%%%%%%%%%%%%%%%%%%%%%%%%%%%%%%%%
\begin{acknowledgments}
\paragraph{Acknowledgments.}
%%%%%%%%%%%%%%%%%%%%%%%%%%%%%%%%%%%%%%%%%%%%%%%%%%%%%%%%%%%%%%%%%%%%%%%
%
We thank Igor Boettcher, Santanu Dey, Igor Herbut, and Joseph Maciejko for illuminating discussions, and Igor Boettcher and Igor Herbut for helpful comments on the manuscript.
This work has been supported by the Deutsche Forschungsgemeinschaft (DFG) through SFB 1143 (A07, Project No.~247310070), the W\"{u}rzburg-Dresden Cluster of Excellence {\it ct.qmat} (EXC 2147, Project No.~390858490), and the Emmy Noether program (JA2306/4-1, Project No.~411750675).
\end{acknowledgments}

%%%%%%%%%%%%%%%%%%%%%%%%%%%%%%%%%%%%%%%%%%%%%%%%%%%%%%%%%%%%%%%%%%%%%%%
% BIBLIOGRAPHY: FOR USE WITH BIBTEX
%%%%%%%%%%%%%%%%%%%%%%%%%%%%%%%%%%%%%%%%%%%%%%%%%%%%%%%%%%%%%%%%%%%%%%%
\bibliographystyle{longapsrev4-2}
\bibliography{QBT3D-Weyl}
%%%%%%%%%%%%%%%%%%%%%%%%%%%%%%%%%%%%%%%%%%%%%%%%%%%%%%%%%%%%%%%%%%%%%%%

%%%%%%%%%%%%%%%%%%%%%%%%%%%%%%%%%%%%%%%%%%%%%%%%%%%%%%%%%%%%%%%%%%%%%%%

\end{document}

\title{Supplemental Material for\\ ``Quasiuniversality from all-in-all-out Weyl quantum criticality in pyrochlore iridates''}

\author{David J. Moser}

\author{Lukas Janssen}

\affiliation{Institut f\"ur Theoretische Physik and W\"urzburg-Dresden Cluster of Excellence ct.qmat, TU Dresden, 01062 Dresden, Germany}

\begin{abstract}
    The Supplemental Material contains 
    (a)~technical details of the mean-field analysis,
    (b)~the derivation of the flow equations, 
    (c)~definitions and numerical values of the solid-angle integrals, 
    (d)~a discussion of emergent particle-hole symmetry at the quantum critical point,
    (e)~a review of the unstable fixed points,
    (f)~an investigation of higher-loop corrections,
    and 
    (g)~a comparison of the quasiuniversal flow with a generic flow.
\end{abstract}

\date{February 20, 2024}
%%% REPLACE "\today" with arXiv submission date %%%

\maketitle

%%%%%%%%%%%%%%%%%%%%%%%%%%%%%%%%%%%%%%%%%%%%%%%%%%%%%%%%%%%%%%%%%%%%%%%
\section{Mean-field theory}
%%%%%%%%%%%%%%%%%%%%%%%%%%%%%%%%%%%%%%%%%%%%%%%%%%%%%%%%%%%%%%%%%%%%%%%
%
In this supplemental section, we discuss technical details of the mean-field analysis.
Formally, the mean-field approximation can be understood as the limit of large number $N$ of quadratic band touching points at the Fermi level. In this limit, fluctuations of the order-parameter field are suppressed.
The mean-field energy is then obtained by integrating out the fermions, leading to the usual logarithm of the fermion determinant in the effective action for the order parameter. Performing the frequency integration leads to the mean-field energy
\begin{align}
    E_\text{MF} (\phi) = \frac{r}{2} \phi^2 + \int_0^\Lambda \frac{\dd[3]{p}}{(2 \pi)^3} \left[ \varepsilon_{\phi}^{(1)} (\vec{p}) + \varepsilon_{\phi}^{(2)} (\vec{p}) \right],
\end{align}
in accordance with Refs.~\cite{janssen15, ray21}.
While the first summand penalizes the presence of a finite order parameter, the second one accounts for a lowering of the total energy by AIAO ordering, arising from the reduction of the low-energy density of states.
The momentum integration is carried out up to an ultraviolet cutoff $\Lambda$, and is performed over the two filled fermionic bands
\begin{multline}
    \varepsilon_\phi^{(1,2)} (\vec{p}) = -\left[(1-\delta)^2 p^4 + 4 \delta \sum_{i=1}^3 d_i^2 (\vec{p}) + (g \phi)^2 \right.
\nonumber\\
    \left. \pm 2 (1+\delta) |g \phi| \sqrt{\sum_{i=1}^3 d_i^2 (\vec{p})} \right]^{1/2},
\end{multline}
which are obtained by diagonalizing the mean-field Hamiltonian $H_\text{MF} = \sum_{a = 1}^5 (1 + s_a \delta) d_a(\vec{p}) \gamma_a + g \phi \gamma_{45}$. Due to the non-trivial anisotropy dependence of the integrand, this integral has to be performed numerically. Minimizing the resulting mean-field energy yields the phase diagram displayed in Fig.~{\figMFT} of the main text.

%%%%%%%%%%%%%%%%%%%%%%%%%%%%%%%%%%%%%%%%%%%%%%%%%%%%%%%%%%%%%%%%%%%%%%%
\section{Derivation of flow equations}
%%%%%%%%%%%%%%%%%%%%%%%%%%%%%%%%%%%%%%%%%%%%%%%%%%%%%%%%%%%%%%%%%%%%%%%

%%%%%%%%%%%%%%%%%%%%%%%%%%%%%%%%%%%%%%%%%%%%%%%%%%%%%%%%%%%%%%%%%%%%%%%
\begin{figure}[tb]
\centering
\includegraphics[width=\linewidth]{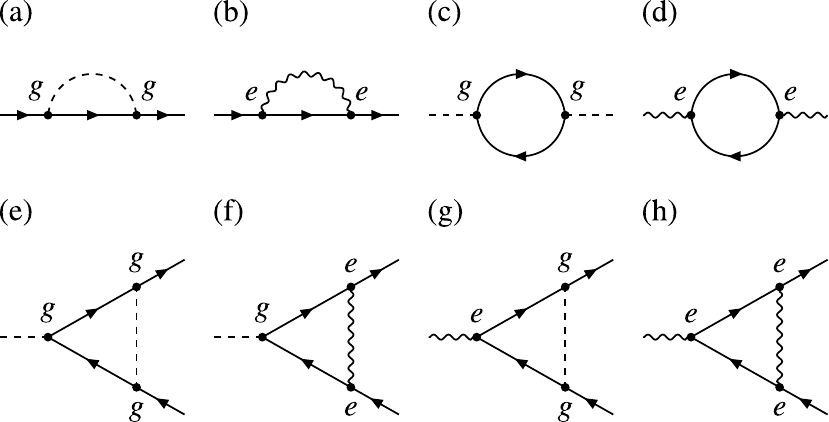}
\caption{Feynman diagrams at the one-loop order contributing to 
(a,b) the fermion anomalous dimensions $\eta_1$, $\eta_\psi$ and the anisotropy parameter renormalization $\Delta\delta$,
(c) the order-parameter anomalous dimension $\eta_\phi$ and the tuning parameter renormalization $\Delta r$,
(d) the Coulomb anomalous dimension $\eta_a$,
and (e)--(h) the vertex renormalizations $\Delta g$ and $\Delta e$, respectively.
The contributions to the flow of $e^2$ from (a) and (g), as well as those from (b) and (h), cancel as a consequence of a Ward identity.}
\label{FigDiagrams}
\end{figure}
%%%%%%%%%%%%%%%%%%%%%%%%%%%%%%%%%%%%%%%%%%%%%%%%%%%%%%%%%%%%%%%%%%%%%%%

In this supplemental section, we give details on the derivation of the flow equations. 
In order to generalize our theory to noninteger spatial dimensions $2<d<4$, we keep the general counting of dimensions in the couplings, but perform the angular integrations directly in the physical dimension $d=3$~\cite{vojta00}. This prescription is believed to be the most appropriate choice for the present type of models~\cite{janssen15}.
Integrating out fast modes with momenta $q$ in the thin shell $\Lambda / b < q < \Lambda$ and arbitrary frequencies $\omega \in \mathbb{R}$ leads to the renormalized action for the slow modes as
\begin{align}
    S_< & = \int\limits_{-\infty}^\infty \frac{\dd{\omega}}{2 \pi} \int\limits_0^{\Lambda/b} \frac{\dd[d]{q}}{(2 \pi)^d} \Bigg\{ 
    \sum_{i=1}^N \psi_i^\dagger \bigg[ 
    b^{\eta_1} \rmi \omega 
    \nonumber\\&\quad
    + b^{\eta_\psi} \sum_{a=1}^5 d_a(\vec{q}) \gamma_a + (\delta + \Delta \delta) \sum_{a=1}^5 s_a d_a(\vec{q}) \gamma_a \bigg] \, \psi_i
    \nonumber\\&\quad
    + \frac{1}{2} \phi \left[ b^{\eta_\phi} q^2 + (r+\Delta r)\right] \phi 
    + \frac{1}{2} a b^{\eta_a} q^2 a 
    \Bigg\} \nonumber\\&\quad
    +\int\limits_{-\infty}^\infty \frac{\dd{\omega_1} \dd{\omega_2}}{(2 \pi)^2} \int\limits_0^{\Lambda/b} \frac{\dd[d]{q_1} \dd[d]{q_2}}{(2 \pi)^{2d}} \Bigg[
    (g + \Delta g) \phi \sum_{i=1}^N \psi_i^\dagger \gamma_{45} \psi_i
    \nonumber\\&\quad
    + \rmi (e + \Delta e) a \sum_{i=1}^N \psi_i^\dagger \psi_i
    \Bigg]\,.
\end{align}
Figure~\ref{FigDiagrams} shows the pertinent diagrams at the one-loop order, giving rise to the anomalous dimensions $\eta_1$, $\eta_\psi$, $\eta_\phi$, $\eta_a$, the explicit anisotropy parameter renormalization $\Delta \delta$, and the vertex renormalizations $\Delta g$, $\Delta e$.
We rescale all frequencies as $b^z \omega \mapsto \omega$, with dynamical exponent $z = 2 + \eta_1 - \eta_\psi$, and all momenta as $b \vec{q} \mapsto \vec{q}$.
Further, we renormalize the fields according to $b^{-(2+d+z-\eta_\psi)/2} \psi \mapsto \psi$, $b^{-(2+d+z-\eta_\phi)/2} \phi \mapsto \phi$, and $b^{-(2+d+z-\eta_a)/2} a \mapsto a$.
The effective action then becomes of the same form as the original action, but with renormalized parameters $\delta$, $g$, $e$, and $r$.
Performing the loop integration leads to the flow equations for $r=0$ as given in Eqs.~\eqbetaFunctionDelta--\eqbetaFunctionE\ of the main text.

In particular, the one-loop fermion self-energy diagrams in Figs.~\ref{FigDiagrams}(a) and (b) are frequency independent, such that $\eta_1 = 0$.
This leads to the scaling relation $z = 2 - \eta_\psi$, 
with the fermion anomalous dimension
\begin{align} \label{eq:etapsi}
    \eta_\psi &= \frac{2}{15} (1 - \delta^2)
    \big[
    (1 - \delta) (g^2 + e^2) f_{1 \text{e}}
    %
    %\nonumber\\&\quad
    %
    - (1 + \delta) (g^2 - e^2)  f_{1 \text{t}}
    \big]\,,
\end{align}
%
%\begin{align}
%    \eta_\psi &= \frac{2}{15} (1 - \delta^2) \big\{[(1 - \delta) f_{1 \text{e}} - (1 + \delta) f_{1 \text{t}}]\,g^2
%    %
%    \nonumber\\&\quad
%    %
%    + [(1 - \delta) f_{1 \text{e}} + (1 + \delta) f_{1 \text{t}}]\,e^2 \big\}\,.
%    %
%\end{align}
%
where we have rescaled the couplings as in Eqs.~\eqbetaFunctionDelta--\eqbetaFunctionE\ of the main text.
Since there appears to be no fundamental reason for the cancellation of frequency dependences in the fermion self-energy diagrams in general, we should expect the above scaling relation to receive corrections at higher loop orders. At the quantum critical point at $\delta_\star = -1$, however, these higher-loop corrections vanish, cf.\ Sec.~\ref{sec:higher-loop}.

By contrast, the self-energy contributions in Figs.~\ref{FigDiagrams}(a) and (b) to the flow of $e^2$ cancel with the explicit charge vertex renormalizations in Figs.~\ref{FigDiagrams}(g) and (h), respectively, as a consequence of the Ward identity associated with the gauge transformation $\psi \mapsto \rme^{\rmi e \lambda(\tau)}\psi$, $a \mapsto a - \partial_\tau \lambda(\tau)$~\cite{janssen17}.
The flow of the effective charge can therefore be written as
\begin{align} \label{eq:betaFunctionE2_supp}
    \frac{\rmd e^2}{\rmd \ln b} = \left(\epsilon + z - 2 - \eta_a + \frac{2\delta}{1-\delta^2}\frac{\rmd \delta}{\rmd \ln b}\right) e^2,
\end{align}
where the last term arises from the reparametrization of the couplings as given as given below Eqs.~\eqbetaFunctionDelta--\eqbetaFunctionG\ in the main text. Equation~\eqref{eq:betaFunctionE2_supp} leads to the scaling relation $\eta_a = \epsilon + z - 2$ at any charged fixed point in the Luttinger fermion system.
As this scaling relation ultimately arises from a Ward identity, which holds at all loop orders within the $\epsilon$ expansion, we expect it to be exact also at a quantum critical point away from $\delta_\star = -1$.

Finally, the flow equation for the tuning parameter $r$ reads
\begin{align}
    \frac{\rmd r}{\rmd \ln b} = (2 - \eta_\phi) r - \frac{4N}{5} (1 - \delta) (1 - \delta^2) g^2 f_{2 \text{e}}\,,
\end{align}
where we have rescaled $r \Lambda^{-2} \mapsto r$, and $g^2$ and $e^2$ as in Eqs.~\eqbetaFunctionDelta--\eqbetaFunctionE\ of the main text. At the critical point, it determines the correlation-length exponent, leading to another scaling relation $1/\nu = 2 - \eta_\phi$. This scaling relation is basically a consequence of the fact that all bosonic selfinteractions compatible with the symmetries of the model, such as a $\phi^4$ term, are irrelevant in the renormalization group sense. We therefore expect it also to hold to all loop orders.

%%%%%%%%%%%%%%%%%%%%%%%%%%%%%%%%%%%%%%%%%%%%%%%%%%%%%%%%%%%%%%%%%%%%%%%
\section{Solid-angle integrals}
%%%%%%%%%%%%%%%%%%%%%%%%%%%%%%%%%%%%%%%%%%%%%%%%%%%%%%%%%%%%%%%%%%%%%%%

%%%%%%%%%%%%%%%%%%%%%%%%%%%%%%%%%%%%%%%%%%%%%%%%%%%%%%%%%%%%%%%%%%%%%%%
\begin{table*}[bt!]
    \caption{Values of solid-angle integrals $f_i(\delta)$ for limiting cases $\delta = \pm 1$ and isotropic case $\delta = 0$ from, whenever possible, analytical integration, otherwise determined numerically.}
    \begin{subtable}[c]{0.25\textwidth}
    \begin{tabular*}{\linewidth}{@{\extracolsep{\fill} } c c c c }
    \hline\hline
        $f_i$ & $f_i (-1)$ & $f_i(0)$ & $f_i (+1)$ \\ \hline
        $f_{1}$ & 1.0942 & 1 & 0.8130 \\
        $f_{1 \text{t}}$ & 1.3294 & 1 & 0.6225 \\
        $f_{1 \text{e}}$ & 0.7415 & 1 & 1.0987 \\
        $f_{2}$ & 1 & 1 & 1/2 \\
        $f_{2 \text{t}}$ & 5/6 & 1 & 0.6775 \\
        $f_{2 \text{e}}$ & 1.3678 & 1 & 5/8 \\
        $f_{3}$ & 4/3 & 1 & 2/3 \\
    \hline\hline
    \end{tabular*}
    \end{subtable}
    \hfill
    \begin{subtable}[c]{0.25\textwidth}
    \centering
    \begin{tabular*}{\linewidth}{@{\extracolsep{\fill} } c c c c }
    \hline\hline
        $f_i$ & $f_i (-1)$ & $f_i(0)$ & $f_i (+1)$ \\ \hline
        $f_{3 \text{t}}$ & 5/9 & 1 & 5/9 \\
        $f_{3 \text{e}}$ & 5/3 & 1 & 5/12 \\
        $f_{3 \overline{\text{t}}}$ & 35/54 & 1 & 0.7262 \\
        $f_{4}$ & 16/5 & 1 & 8/5 \\
        $f_{4 \text{t}}$ & 4/3 & 1 & 4/9 \\
        $f_{4 \text{e}}$ & 4/3 & 1 & 1 \\
        $f_{4 \text{t} \text{t}}$ & 7/27 & 1 & 7/9 \\
    \hline\hline
    \end{tabular*}
    \end{subtable}
    \hfill
    \begin{subtable}[c]{0.25\textwidth}
    \begin{tabular*}{\linewidth}{@{\extracolsep{\fill} } c c c c }
    \hline\hline
        $f_i$ & $f_i (-1)$ & $f_i(0)$ & $f_i (+1)$ \\ \hline
        $f_{4 \text{t} \text{t}'}$ & 7/9 & 1 & 7/18 \\
        $f_{4 \text{e} \text{e}}$ & 7/3 & 1 & 7/16 \\
        $f_{4 \text{e} \text{t}}$ & 7/9 & 1 & 7/24 \\
        $f_{4 \text{t} \overline{\text{t}}}$ & 77/162 & 1 & 0.8876 \\
        $f_{4 \text{e} \overline{\text{t}}}$ & 77/108 & 1 & 77/144 \\
        $f_{e^2}$ & 16/9 & 1 & 4/3 \\
        $f_{g^2}$ & 16/9 & 0 & $-2/3$ \\
    \hline\hline
    \end{tabular*}
    \end{subtable}
    \label{TablefiLimits}
\end{table*}
%%%%%%%%%%%%%%%%%%%%%%%%%%%%%%%%%%%%%%%%%%%%%%%%%%%%%%%%%%%%%%%%%%%%%%%

%%%%%%%%%%%%%%%%%%%%%%%%%%%%%%%%%%%%%%%%%%%%%%%%%%%%%%%%%%%%%%%%%%%%%%%
\begin{figure*}[tb!]
\includegraphics[width=\linewidth]{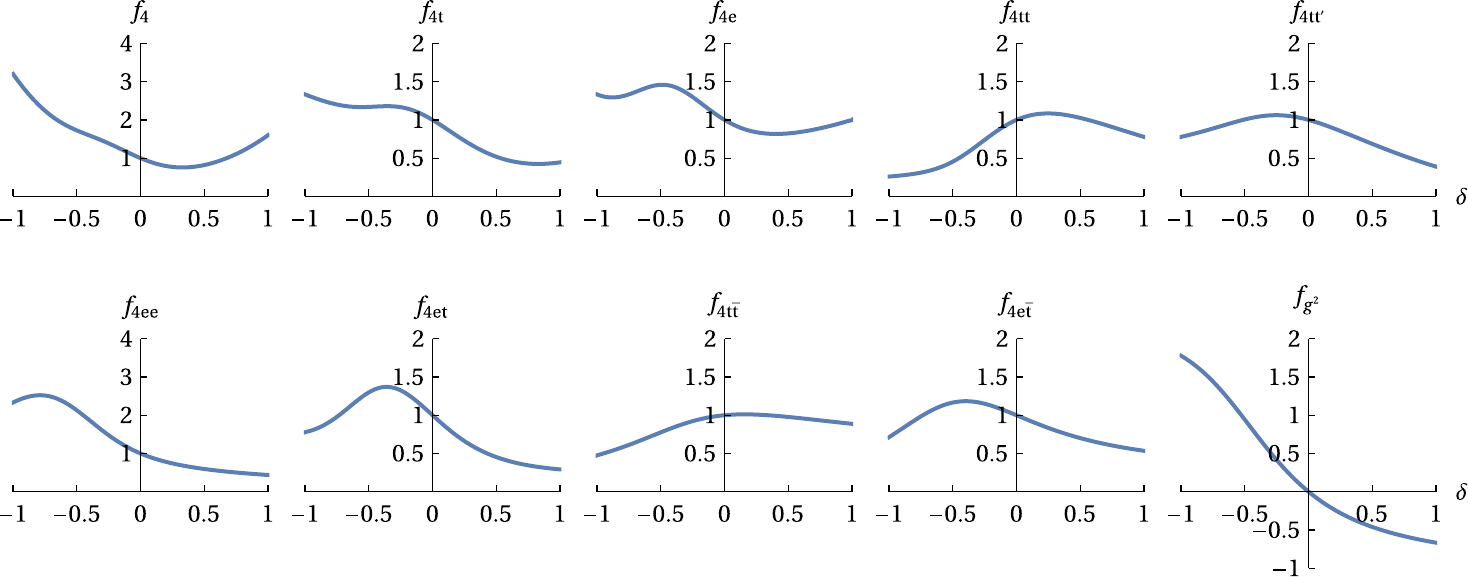}
\caption{Graphs of solid-angle integrals $f_i$ as function of $\delta$. Except for $f_{g^2}$, all solid-angle integrals satisfy $f_i > 0$ for $\delta \in [-1,1]$ and $f_i = 1$ for $\delta = 0$. Graphs not shown here are given in Ref.~\cite{boettcher17}.}
\label{fig:fi}
\end{figure*}
%%%%%%%%%%%%%%%%%%%%%%%%%%%%%%%%%%%%%%%%%%%%%%%%%%%%%%%%%%%%%%%%%%%%%%%

In this supplemental section, we provide definitions and numerical values of the solid-angle integrals $f_i \equiv f_i(\delta)$, with $i \in \{1$, $1\text{t}$, $1\text{e}$, $2$, $2\text{t}$, $2\text{e}$, $3$, $3\text{t}$, $3\text{e}$, $3\overline{\text{t}}$, $4$, $4\text{t}$, $4\text{e}$, $4\text{tt}$, $4\text{tt}'$, $4\text{ee}$, $4\text{et}$, $4\text{t}\overline{\text{t}}$, $4\text{e}\overline{\text{t}}, e^2, g^2\}$, as function of the anisotropy parameter $\delta$, occurring in the loop expansion.
These are bounded and continuous functions of order unity with $f_i >0$ ($f_{g^2} \geq -2/3$) for $\delta \in [-1,1]$ and $f_i = 1$ ($f_{g^2} = 0$) for $\delta =0$.
Some of these have already been defined in Ref.~\cite{boettcher17}, reading
\begin{align}
    f_{1} (\delta) &\coloneqq \frac{1}{4 \pi} \int \dd{\Omega} \frac{1}{\tilde{X}^{1/2}}, \allowdisplaybreaks[1] \\
    f_{1 \text{t}} (\delta) &\coloneqq \frac{5}{4 \pi} \int \dd{\Omega} \frac{\tilde{d}_1^2}{\tilde{X}^{1/2}}, \allowdisplaybreaks[1] \\
    f_{1 \text{e}} (\delta) &\coloneqq \frac{5}{4 \pi} \int \dd{\Omega} \frac{\tilde{d}_4^2}{\tilde{X}^{1/2}}, \allowdisplaybreaks[1] \\
    f_{2} (\delta) &\coloneqq \frac{1}{4 \pi} (1-\delta) (1+\delta) \int \dd{\Omega} \frac{1}{\tilde{X}^{3/2}}, \allowdisplaybreaks[1] \\
    f_{2 \text{t}} (\delta) &\coloneqq \frac{5}{4 \pi} (1+\delta) \int \dd{\Omega} \frac{\tilde{d}_1^2}{\tilde{X}^{3/2}}, \allowdisplaybreaks[1] \\
    f_{2 \text{e}} (\delta) &\coloneqq \frac{5}{4 \pi} (1-\delta) \int \dd{\Omega} \frac{\tilde{d}_4^2}{\tilde{X}^{3/2}}, \allowdisplaybreaks[1] \\
    f_{3} (\delta) &\coloneqq \frac{1}{4 \pi} (1-\delta)^3 (1+\delta)^3 \int \dd{\Omega} \frac{1}{\tilde{X}^{5/2}}, \allowdisplaybreaks[1] \\
    f_{3 \text{t}} (\delta) &\coloneqq \frac{5}{4 \pi} (1-\delta) (1+\delta)^3 \int \dd{\Omega} \frac{\tilde{d}_1^2}{\tilde{X}^{5/2}}, \allowdisplaybreaks[1] \\
    f_{3 \text{e}} (\delta) &\coloneqq \frac{5}{4 \pi} (1-\delta)^3 (1+\delta) \int \dd{\Omega} \frac{\tilde{d}_4^2}{\tilde{X}^{5/2}}, \allowdisplaybreaks[1] \\
    f_{3 \overline{\text{t}}} (\delta) &\coloneqq \frac{35}{\sqrt{3} 4 \pi} (1+\delta)^3 \int \dd{\Omega} \frac{\tilde{d}_1 \tilde{d}_2 \tilde{d}_3}{\tilde{X}^{5/2}},
\end{align}
where $\int \dd{\Omega} \coloneqq \int_0^\pi \dd{\theta} \sin \theta \int_0^{2 \pi} \dd{\phi}$ denotes the integration over the solid angle, $\tilde{X} (\theta, \phi) \coloneqq (1-\delta)^2 + 12 \delta \sum_{i<j} \frac{q_i^2}{q^2} \frac{q_j^2}{q^2}$, and $\tilde{d}_a (\theta, \phi) \coloneqq d_a (\vec{q}) / q^2$ are the $\ell = 2$ real spherical harmonics.
Upon spatial rotations, the latter transform under the irreducible representation $\text{T}_{2 \text{g}}$ ($\text{E}_\text{g}$) of the octahedral point group $\text{O}_\text{h}$ for $a=1,2,3$ ($a=4,5$), and the indices t and e of the functions $f_i$ indicate the type of spherical harmonics involved in the integral.
In addition to the above-defined functions, the loop expansion of the order-parameter field theory  [Eqs.~\EqLagrangianFree--\EqLagrangianOP\ of the main text] gives rise to the following new solid-angle integrals,
\begin{align}
    f_{4} (\delta) &\coloneqq \frac{1}{4 \pi} (1-\delta)^5 (1+\delta)^5 \int \dd{\Omega} \frac{1}{\tilde{X}^{7/2}},\\
    f_{4 \text{t}} (\delta) &\coloneqq \frac{5}{4 \pi} (1-\delta)^3 (1+\delta)^5 \int \dd{\Omega} \frac{\tilde{d}_1^2}{\tilde{X}^{7/2}}, \allowdisplaybreaks[1] \\
    f_{4 \text{e}} (\delta) &\coloneqq \frac{5}{4 \pi} (1-\delta)^5 (1+\delta)^3 \int \dd{\Omega} \frac{\tilde{d}_4^2}{\tilde{X}^{7/2}}, \allowdisplaybreaks[1] \\
    f_{4 \text{t} \text{t}} (\delta) &\coloneqq \frac{35}{12 \pi} (1-\delta) (1+\delta)^5 \int \dd{\Omega} \frac{\tilde{d}_1^2 \cdot \tilde{d}_1^2}{\tilde{X}^{7/2}}, \allowdisplaybreaks[1] \\
    f_{4 \text{t} \text{t}'} (\delta) &\coloneqq \frac{35}{4 \pi} (1-\delta) (1+\delta)^5 \int \dd{\Omega} \frac{\tilde{d}_1^2 \cdot \tilde{d}_2^2}{\tilde{X}^{7/2}}, \allowdisplaybreaks[1] \\
    f_{4 \text{e} \text{e}} (\delta) &\coloneqq \frac{35}{12 \pi} (1-\delta)^5 (1+\delta) \int \dd{\Omega} \frac{\tilde{d}_4^2 \cdot \tilde{d}_4^2}{\tilde{X}^{7/2}}, \allowdisplaybreaks[1] \\
    f_{4 \text{e} \text{t}} (\delta) &\coloneqq \frac{35}{4 \pi} (1-\delta)^3 (1+\delta)^3 \int \dd{\Omega} \frac{\tilde{d}_1^2 \cdot \tilde{d}_4^2}{\tilde{X}^{7/2}}, \allowdisplaybreaks[1] \\
    f_{4 \text{t} \overline{\text{t}}} (\delta) &\coloneqq \frac{385}{\sqrt{3} 12 \pi} (1+\delta)^5 \int \dd{\Omega} \frac{\tilde{d}_1^2 \cdot \tilde{d}_1 \tilde{d}_2 \tilde{d}_3}{\tilde{X}^{7/2}}, \allowdisplaybreaks[1] \\
    f_{4 \text{e} \overline{\text{t}}} (\delta) &\coloneqq \frac{385}{\sqrt{3} 4 \pi} (1-\delta) (1+\delta)^3 \int \dd{\Omega} \frac{\tilde{d}_4^2 \cdot \tilde{d}_1 \tilde{d}_2 \tilde{d}_3}{\tilde{X}^{7/2}}.
\end{align}
For the Coulomb and order-parameter anomalous dimensions, it is convenient to define two further functions as combinations of the above-defined ones,
\begin{align}
    f_{e^2} (\delta) \coloneqq& \frac{1}{3} \left[ 2 (1 - \delta)^2 + 3 (1 + \delta)^2 \right] f_2 (\delta)
    \nonumber\\&
    - \frac{2}{3} \bigg[ \frac{2}{5} (1 - \delta)^2 f_{3 \text{e}} (\delta) + \frac{3}{5} (1 + \delta)^2 f_{3 \text{t}} (\delta)
    \nonumber\\&
    - \frac{12}{5} \frac{\delta^2}{(1 + \delta)^2} f_{3 \text{t}} (\delta) + \frac{36}{35} \frac{\delta^2 (1 - \delta)}{(1 + \delta)^2} f_{3 \overline{\text{t}}} (\delta) \bigg],
    \allowdisplaybreaks[1] \\
    f_{g^2} (\delta) \coloneqq& \frac{2 (1 - \delta)^2 - (1 + \delta)^2}{4} f_2(\delta)
    \nonumber\\&
    + \frac{8 \delta^2 - 6 (1 - \delta)^2 (1 + \delta)^2 - 11 (1 + \delta)^4}{20 (1 + \delta)^2} f_{3 \text{t}}(\delta)\nonumber\\&
    + \frac{4 (1 - \delta)^2 + 9 (1 + \delta)^2}{30} f_{3 \text{e}}(\delta) - \frac{6 \delta^2 (1 - \delta)}{35 (1 + \delta)^2} f_{3 \overline{\text{t}}}(\delta)
    \allowdisplaybreaks[1] \nonumber\\&
    + \Bigg\{ \frac{(1 - \delta)^2 \left[-4 (1 - \delta)^2 + 7 (1 + \delta)^2\right]}{14 (1 + \delta)^2}
    \nonumber\\&\qquad
    + \frac{2 \delta^2 \left[-(1 - \delta)^2 + 9 (1 + \delta)^2\right]}{7 (1 + \delta)^4} \Bigg\} f_{4 \text{t} \text{t}}(\delta)
    \allowdisplaybreaks[1] \nonumber\\&
    + \Bigg\{ \frac{(1 - \delta)^2 \left[-5 (1 - \delta)^2 + 
   11 (1 + \delta)^2\right]}{42 (1 + \delta)^2}
   \nonumber\\&\qquad
    + \frac{2 \delta^2 \left[(1 - \delta)^2 + 18 (1 + \delta)^2\right]}{21 (1 + \delta)^4} \Bigg\} f_{4 \text{t} \text{t}'}(\delta)
    \allowdisplaybreaks[1] \nonumber\\&
    - \frac{4 (1 - \delta)^2}{21}  f_{4 \text{e} \text{e}}(\delta) + \frac{54 \delta^2 (1 - \delta)}{77 (1 + \delta)^2} f_{4 \text{t} \overline{\text{t}}}(\delta)
    \allowdisplaybreaks[1] \nonumber\\&
    - \frac{4 \delta \left[(1 - \delta)^2 + 3 (1 + \delta)^2\right]}{21 (1 + \delta)^2} f_{4 \text{e} \text{t}}(\delta)
    \allowdisplaybreaks[1] \nonumber\\&
    - \frac{12 \delta^2 (1 - \delta)^2}{77 (1 + \delta)^2} f_{4 \text{e} \overline{\text{t}}}(\delta).
\end{align}
We emphasize that all $f_i$ are bounded from above and below for all $\delta \in [-1,1]$; in particular, they remain finite in the limiting cases $\delta = \pm 1$, see Table~\ref{TablefiLimits}.
Figure~\ref{fig:fi} shows the graphs of those functions $f_i$ that have not already been defined in Ref.~\cite{boettcher17}.

%%%%%%%%%%%%%%%%%%%%%%%%%%%%%%%%%%%%%%%%%%%%%%%%%%%%%%%%%%%%%%%%%%%%%%%
\section{Emergent particle-hole symmetry}
%%%%%%%%%%%%%%%%%%%%%%%%%%%%%%%%%%%%%%%%%%%%%%%%%%%%%%%%%%%%%%%%%%%%%%%

%%%%%%%%%%%%%%%%%%%%%%%%%%%%%%%%%%%%%%%%%%%%%%%%%%%%%%%%%%%%%%%%%%%%%%%
\begin{table*}[t!]
    \caption{Fixed points, their locations, and number of relevant directions on the critical hypersurface $r=0$.}
    \centering
    \begin{tabular*}{\linewidth}{@{\extracolsep{\fill} } l c c c c l}
    \hline\hline
        Label & $g^2 N/\epsilon$ & $e^2 N/\epsilon$ & $\delta$ &  Relevant directions & Comment \\ \hline
        G & 0 & 0 & $\in [-1,1]$ & 2 & Line of Gaussian fixed points \\
        A & 0 & $9 / 16$ & $-1$ & 2 & Unstable fixed point at $\delta = -1$, observed in~\cite{boettcher17} \\
        LAB & 0 & $15 / 19$ & 0 & 1 & Luttinger-Abrikosov-Bene\-slavs\-kii fixed  point, discussed in~\cite{abrikosov71,abrikosov74,moon13,herbut14} \\
        A' &0 & $3 / 4$ & 1 & 2 & Unstable fixed point at $\delta = 1$, observed in~\cite{boettcher17} \\
        QCP$_0$ & $9 / 16$ & 0 & $-1$ & 1 & Quantum critical fixed point for uncharged theory $e^2 = 0$\\
        QCP &$9 / 16$ & $9 / 16$ & $-1$ & 0 & Unique stable fixed point for AIAO Weyl quantum criticality\\
    \hline\hline
    \end{tabular*}
    \label{TableFixedPoints}
\end{table*}
%%%%%%%%%%%%%%%%%%%%%%%%%%%%%%%%%%%%%%%%%%%%%%%%%%%%%%%%%%%%%%%%%%%%%%%

In this supplemental section, we demonstrate that particle-hole symmetry is emergent at the quantum critical point.
To this end, we add to the Lagrangian the perturbation $L_x = -x \sum_{i=1}^N \psi^\dagger_i \nabla^2 \psi_i$, with small parameter $x$, $|x| \ll 1$.
For $x \neq 0$, $L_x$ breaks particle-hole symmetry explicitly.
The corresponding flow equation reads
\begin{align}
    \frac{\rmd x}{\rmd \ln b} = - \eta_\psi x,
\end{align}
with $\eta_\psi$ given in Eq.~\eqref{eq:etapsi}.
Importantly, $\eta_\psi>0$ ($\eta_\psi=0$) for $g^2>0$, $e^2>0$, and $\delta \in (-1,0]$ ($\delta = -1$), corresponding to an irrelevant (marginal) particle-hole asymmetry parameter $x$.
Assuming $\delta < 0$ for the pyrochlore iridates~\cite{kondo15}, $x$ flows to zero towards the infrared, and particle-hole symmetry becomes emergent in the quasiuniversal intermediate-temperature regime.
Note that $x$ is exactly marginal at the stable quantum critical fixed point. There is therefore a line of fixed points at $\delta = -1$ within a finite interval around $x=0$. We have verified, however, that the flow for any small perturbation $x \neq 0$ and $\delta - \delta_\star > 0$ is, in the ultra-low-energy limit, always towards the particle-hole-symmetric quantum critical fixed point at $x=0$.

%%%%%%%%%%%%%%%%%%%%%%%%%%%%%%%%%%%%%%%%%%%%%%%%%%%%%%%%%%%%%%%%%%%%%%%
\section{Fixed-point structure}
%%%%%%%%%%%%%%%%%%%%%%%%%%%%%%%%%%%%%%%%%%%%%%%%%%%%%%%%%%%%%%%%%%%%%%%

In this supplemental section, we give more details of the fixed-point structure on the critical hypersurface $r=0$, including also the unstable fixed points with infrared relevant directions.

First of all, vanishing couplings $g^2 = e^2 = 0$ lead to zero renormalization group flow for arbitrary values of the anisotropy parameter $\delta \in [-1,1]$, leading to a line of Gaussian fixed points [thick black line in Fig.~\figFlowDiagrams(a) of the main text]. 

For $g^2 = 0$, but $e^2 > 0$, the flow equations host three interacting fixed points. One, located at $\delta = 0$, corresponds to the well-known Luttinger-Abrikosov-Beneslavskii fixed point~\cite{abrikosov71, abrikosov74, moon13, herbut14}. It is stable within the plane $g^2 = 0$ [see Fig.~\figFlowDiagrams(d) of the main text], but unstable in the direction perpendicular to it, as long as $r$ is tuned to criticality [Fig.~\figFlowDiagrams(c) of the main text].
Two further unstable fixed points are located at $\delta = \pm 1$ and $g^2=0$, both of which have previously been encountered~\cite{boettcher17}, but are of no great importance to this work, since not only are they unstable within the $g^2=0$ plane, but also in the direction perpendicular to it.

For $e^2 = 0$, but $g^2 > 0$, we find an additional interacting fixed point, corresponding to AIAO quantum criticality in the absence of the long-range Coulomb repulsion, labeled as QCP$_0$ in Figs.~\figFlowDiagrams(a) and \figFlowDiagrams(b) of the main text.
It is stable within the $e^2=0$ plane, but unstable in the direction perpendicular to it.

%%%%%%%%%%%%%%%%%%%%%%%%%%%%%%%%%%%%%%%%%%%%%%%%%%%%%%%%%%%%%%%%%%%%%%%
\begin{figure}[b]
    \centering
    \includegraphics[scale=1.25]{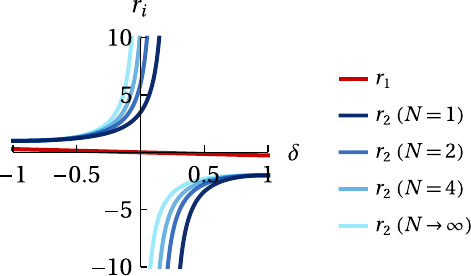}
    \caption{Right-hand sides of Eqs.~\eqref{eq:r1} and \eqref{eq:r2} as function of $\delta$, showing that $r_1 \neq r_2$ for all values of $\delta \in [-1,1]$, preventing the existence of any interacting fixed point at $\delta \neq \pm 1$ for an arbitrary number of band touching points $N \in \mathbb{N}$.}
    \label{fig:numArg}
\end{figure}
%%%%%%%%%%%%%%%%%%%%%%%%%%%%%%%%%%%%%%%%%%%%%%%%%%%%%%%%%%%%%%%%%%%%%%%

There is thus a unique stable fixed point on the critical hypersurface $r=0$, which is the quantum critical point discussed in the main text, located at $(\delta_\star, g^2_\star, e^2_\star) = (-1, 9\epsilon/(16N), 9\epsilon/(16N))$, and labeled as QCP in Figs.~\figFlowDiagrams(a)--\figFlowDiagrams(c) of the main text.

The fact that no further interacting real fixed point exists at some other value of $\delta  > -1$ on the critical hypersurface can also be seen as follows:
Assume such a fixed point existed. The fact that $f_{g^2}(\delta = 1) < 0$ implies that such additional fixed point would need to be located at $\delta < 1$, since any real fixed point at $\delta = 1$ is prevented by the fixed-point condition $\rmd g^2/(\rmd\ln b)=0$ as long as $g^2>0$, see Eq.~\eqbetaFunctionG\ of the main text.
For $\delta \neq \pm 1$, however, the fixed-point condition $\rmd \delta/(\rmd\ln b)=0$ can only be satisfied if
\begin{align} \label{eq:r1}
    \frac{g^2}{e^2} = \frac{f_{1 \text{t}}(\delta) - f_{1 \text{e}}(\delta)}{f_{1 \text{t}}(\delta) + f_{1 \text{e}}(\delta)} \eqqcolon r_1(\delta)\,.
\end{align}
A second relation, which simultaneously has to hold at the putative fixed point, arises from the condition that the coupling ratio $(g^2/e^2)$ does not flow, $\rmd (g^2/e^2)/(\rmd\ln b)=0$, which can be simplified to 
\begin{align} \label{eq:r2}
    \frac{g^2}{e^2} = \frac{2 (1 - \delta)^2 (1 + \delta) f_{2 \text{e}}(\delta) + 5 N f_{e^2}(\delta)}{2 (1 - \delta)^2 (1 + \delta) f_{2 \text{e}}(\delta) + 5 N f_{g^2}(\delta)} \eqqcolon r_2(\delta)\,.
\end{align}
The right-hand sides of Eqs.~\eqref{eq:r1} and \eqref{eq:r2} define functions $r_i(\delta)$, $i=1,2$.
Figure~\ref{fig:numArg} shows that $r_1 \neq r_2$ for all values of $\delta \in [-1,1]$ and any number of quadratic band touching points $N \in \mathbb{N}$. This implies that the two fixed-point conditions for a putative interacting fixed point located at $\delta \neq \pm 1$,  Eqs.~\eqref{eq:r1} and \eqref{eq:r2}, cannot be simultaneously satisfied.
At the one-loop order, there are therefore no other interacting fixed points; in particular, the stable quantum critical fixed point discussed in the main text is unique.

All fixed points on the critical hypersurface $r=0$, their locations, and number of infrared relevant directions, are summarized in Table~\ref{TableFixedPoints}.

%%%%%%%%%%%%%%%%%%%%%%%%%%%%%%%%%%%%%%%%%%%%%%%%%%%%%%%%%%%%%%%%%%%%%%%
\section{Higher-loop corrections}
\label{sec:higher-loop}
%%%%%%%%%%%%%%%%%%%%%%%%%%%%%%%%%%%%%%%%%%%%%%%%%%%%%%%%%%%%%%%%%%%%%%%

In this supplemental section, we substantiate our claim that the critical behavior at the antiferromagnetic Weyl quantum critical point is one-loop exact.
In particular, we argue that higher-loop corrections to the critical exponents vanish when $\delta$ approaches the fixed-point value $\delta_\star = -1$.
We first note any closed fermion loop in a given diagram leads to a pole $\propto 1/(1-\delta^2)$, while any loop containing at least one inner boson line leads to only a regular contribution as function of $\delta$.
This implies that the only diagrams that contribute, at a given loop order, to the critical exponents of a quantum critical point at $\delta_\star = -1$ are those that involve the largest possible number of closed fermion loops.

\begin{figure}[b]
\centering
\includegraphics[width=\linewidth]{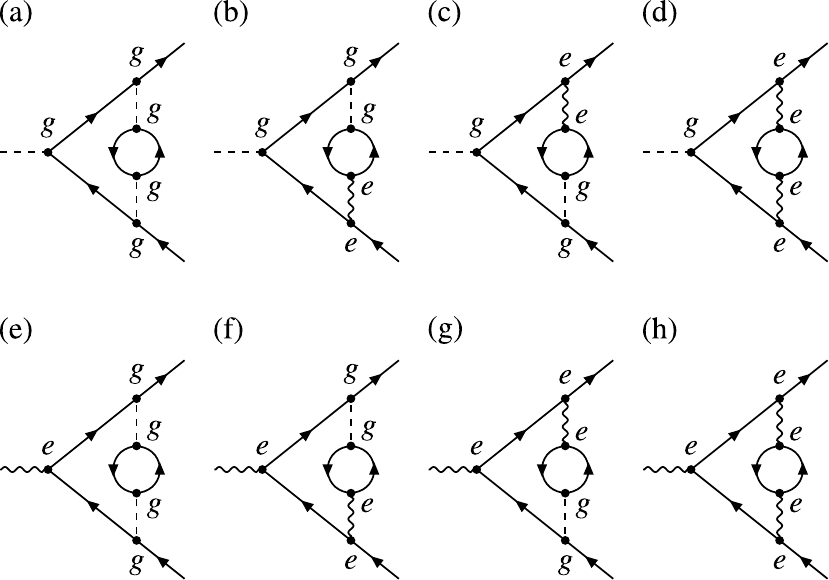}
\caption{Vertex corrections with largest possible number of closed fermion loops at two-loop order, contributing to the renormalization of 
(a)--(d) the Yukawa coupling $g$ and 
(e)--(h) the effective charge~$e$.}
\label{fig:DiagramsTwoLoop}
\end{figure}

To illustrate this point, let us consider the two-loop vertex corrections to the flows of the Yukawa coupling and the effective charge.
The two-loop diagrams with the largest number of closed fermion loops contributing to the explicit vertex corrections are given in Fig.~\ref{fig:DiagramsTwoLoop}.
%
%%%%%%%%%%%%%%%%%%%%%%%%%%%%%%%%%%%%%%%%%%%%%%%%%%%%%%%%%%%%%%%%%%%%%%%
%%%%%%%%%%%%%%%%%%%%%%%%%%%%%%%%%%%%%%%%%%%%%%%%%%%%%%%%%%%%%%%%%%%%%%%
%
These diagrams are most conveniently evaluated using dimensional regularization.
We start with the correction $\Gamma^{(2)}_\phi \propto g^5$ given in Fig.~\ref{fig:DiagramsTwoLoop}(a), which contributes to the flow of the Yukawa coupling. The evaluation of this diagram involves the one-loop order-parameter self-energy at arbitrary external frequencies and momenta,
\begin{align} \label{eq:Sigma_phi}
    \Sigma_\phi^{(1)} (\omega, \vec{p}) = 2 N g^2 \int \frac{\dd[d]{q}}{(2 \pi)^d} \frac{\left(Q_{-}^2+Q_{+}^2\right) \left( Q_\times^4 - Q_{-}^2 Q_{+}^2 \right)}{Q_{-}^2 Q_{+}^2 \left[\omega^2 + \left(Q_{-}^2+Q_{+}^2 \right)^2\right]}.
\end{align}
In the above equation, we have already performed the frequency integral, and have used the abbreviations $Q_{\pm}^4 \equiv \sum_a (1+s_a \delta)^2 d_a^2(\vec{q} \pm \frac{\vec{p}}{2})$ and $Q_\times^4 \equiv \sum_a s_a (1+s_a \delta)^2 d_a(\vec{q} - \frac{\vec{p}}{2}) d_a(\vec{q} + \frac{\vec{p}}{2})$.
In $d=4-\epsilon$ spatial dimensions, the substitution $\vec{q} \mapsto \sqrt{|\omega|} \vec{q}$ reveals the scaling form for the boson self-energy, $\Sigma_\phi^{(1)} (\omega, \vec{p}) = N g^2 |\omega|^{1-\frac{\epsilon}{2}} \mathcal{F}_{\epsilon,\delta}( \vec{p}/\sqrt{|\omega|})$ with scaling function $\mathcal{F}_{\epsilon,\delta}$.
It follows that there is a unique divergent term in the sense of dimensional regularization, which is frequency independent,
\begin{align}
    \Sigma_\phi^{(1)} (\omega, \vec{p}) - \Sigma_\phi^{(1)} (0, 0) = N g^2 \frac{f_{g^2}(\delta)}{1-\delta^2} \frac{p^2}{\epsilon} + \text{finite terms}\,,
\end{align}
displaying the advertised pole $\propto 1/(1-\delta^2)$ as function of $\delta$.
The two-loop correction $\Gamma_\phi^{(2)}$ given in Fig.~\ref{fig:DiagramsTwoLoop}(a) now can be reduced to an expression involving only the one-loop correction $\Gamma_\phi^{(1)}$ given in Fig.~\ref{FigDiagrams}(e), of the form
\begin{align}
    \Gamma_\phi^{(2)} \propto N g^2 \frac{f_{g^2}(\delta)}{1-\delta^2} \Gamma_\phi^{(1)},
\end{align}
with $\delta$-independent prefactors.
We observe that the rescaling of the couplings given below Eqs.~\eqbetaFunctionDelta--\eqbetaFunctionG\ in the main text precisely cancels the $1/(1 - \delta^2)$ divergence caused by the internal fermion loop, such that the two-loop contribution to the flow of the Yukawa coupling reads
\begin{align} \label{eq:g2_two-loop}
    \left.\frac{\rmd g^2}{\rmd \ln b}\right|_\text{Fig.~\ref{fig:DiagramsTwoLoop}(a)} = \frac{2}{5} N (1-\delta) (1-\delta^2) f_\mathrm{2e}(\delta) f_{g^2}(\delta) g^6.
\end{align}
Importantly, for $\delta = -1$, the two-loop correction is suppressed.
Similar scaling relations and reduction formulas, relating two-loop vertex corrections with one-loop boson self-energies and vertex corrections, can be derived for all remaining diagrams shown in Figs.~\ref{fig:DiagramsTwoLoop}(b)--(h).
In particular, the two-loop diagram depicted in Figs.~\ref{fig:DiagramsTwoLoop}(d) gives rise to a completely analogous expression as displayed in Eq.~\eqref{eq:g2_two-loop}.
Moreover, the two-loop contributions in Figs.~\ref{fig:DiagramsTwoLoop}(e,h) turn out to vanish exactly, since the corresponding one-loop diagrams are already zero.
Also, the four remaining two-loop vertex corrections given in Figs.~\ref{fig:DiagramsTwoLoop}(b,c,f,g) are exactly zero, since the closed fermion loop connected to a photon field and an order-parameter field vanishes by symmetry.
In sum, the flows of the Yukawa coupling and the effective charge can be written at the two-loop order as
\begin{align}
    \frac{\rmd g^2}{\rmd \ln b} & = (\epsilon - \eta_\phi) g^2 + \mathcal O(\delta - \delta_\star), 
    \allowdisplaybreaks[1] \\
    \frac{\rmd e^2}{\rmd \ln b} & = (\epsilon - \eta_a) e^2 + \mathcal O(\delta - \delta_\star),
\end{align}
which implies $\eta_\phi = \epsilon + \mathcal O(\epsilon^3)$ and $\eta_a = \epsilon + \mathcal O(\epsilon^3)$ at the quantum critical point. Analogously, from the flow of the tuning parameter $r$, we can show that $1/\nu = 2 - \epsilon +  \mathcal O(\epsilon^3)$ at the two-loop order.
In fact, analogous arguments apply for any given loop order, since the only diagrams that contribute at $\delta = -1$ are those that involve the largest possible number of closed fermion loops, implying that they can be reduced to products of one-loop self-energies and vertex corrections. These, however, will always be suppressed at $\delta = -1$, implying that $\eta_\phi = 4-d$, $\eta_a = 4-d$, $z=2$, and $1/\nu = d - 2$ exactly at the quantum critical point, as stated in the main text.
We emphasize that while these results have been obtained within a particular dimensional continuation scheme, the above analysis of the higher-loop corrections can be carried out in an analogous way (although somewhat less explicitly), without ever specifying the scheme. Our finding that the flow near the critical fixed point is one-loop exact and higher-loop corrections vanish at the quantum critical point, and the resulting critical behavior, is therefore independent of our particular approach, and continues to hold for other dimensional continuation schemes.

%%%%%%%%%%%%%%%%%%%%%%%%%%%%%%%%%%%%%%%%%%%%%%%%%%%%%%%%%%%%%%%%%%%%%%%
\section{Quasiuniversal flow vs.\ generic flow}
\label{sec:quasiuniversal}
%%%%%%%%%%%%%%%%%%%%%%%%%%%%%%%%%%%%%%%%%%%%%%%%%%%%%%%%%%%%%%%%%%%%%%%

%%%%%%%%%%%%%%%%%%%%%%%%%%%%%%%%%%%%%%%%%%%%%%%%%%%%%%%%%%%%%%%%%%%%%%%
\begin{figure}[b]
\centering
\includegraphics[width=\linewidth]{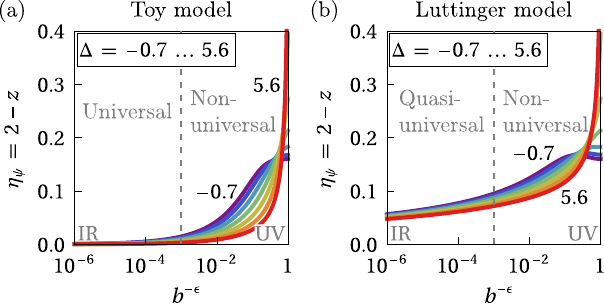}
\caption{%
Flowing fermion anomalous dimension $\eta_\psi$ as function of renormalization group scale $b^{-\epsilon}$ from the ultraviolet scale, corresponding to $b^{-\epsilon} = 1$, to a deep infrared scale, corresponding to $b^{-\epsilon} = 10^{-6}$, starting from different microscopic parameters on the critical hypersurface $r=0$.
Each curve corresponds to a numerical integration of the flow equations using the initial conditions $(\delta, g^2, e^2) = (-0.5, 0.7391, 0.9098 + \Delta)$ for $b^{-\epsilon}=1$, with $\Delta = -0.7, -0.6, -0.4, 0, 0.8, 2.4, 5.6$ from purple to red, as in Fig.~\figquasiunversality\ of the main text.
Panel (a) shows the toy model flow, in which the universal regime emerges at $b^{-\epsilon} \lesssim 10^{-3}$, as indicated by the dashed line, which manifests itself in constant exponents that are independent of any microscopic parameters.
Panel (b) shows the Luttinger model flow for comparison [same as Fig.~\figquasiunversality(c) of the main text], in which the quasiuniversal regime emerges at $b^{-\epsilon} \lesssim 10^{-3}$, as indicated by the dashed line, which manifests itself in slowly drifting exponents.}
\label{fig:QuasiuniversalVsUniversal}
\end{figure}
%%%%%%%%%%%%%%%%%%%%%%%%%%%%%%%%%%%%%%%%%%%%%%%%%%%%%%%%%%%%%%%%%%%%%%%

In this supplemental section, we compare the quasiuniversal flow with a generic power-law flow of a toy model that does not feature a marginal coupling, illustrating the unusual behavior observed in the quasiuniversal regime.
To construct such a toy flow, we manually modify the flow equation of the anisotropy $\delta$ in order to remove its ``marginality.'' To this end, we simply replace $(1+\delta)^2$ by $(1+\delta)$, giving rise to the toy flow equation
\begin{align} \label{eq:betaFunctionDeltaToy}
    \frac{\rmd\delta}{\rmd \ln b} = 
    - \frac{2}{15} (1-\delta)^2 (1+\delta)
    \left[
    (g^2 + e^2) f_{1 \text{e}} + (g^2 - e^2) f_{1 \text{t}}
    \right] .
\end{align}
We leave the flow equations for the Yukawa coupling $g^2$ and the effective charge $e^2$ unchanged.
As a consequence, the system still features an AIAO Weyl quantum critical point at $\delta_\star = -1$, characterized by the same leading universal critical exponents $z=2$, $\eta_\psi = 0$, $\eta_\phi = \eta_a = \epsilon$, and $1/\nu = 2 - \epsilon$.
However, the flow towards this fixed point in this toy model is qualitatively different, since the anisotropy $\delta$ is no longer marginally irrelevant, but now represents a standard power-law irrelevant parameter at the quantum critical point.
This can be seen by expanding the above flow equation about the fixed point, reading
\begin{align}
    \dv{(\delta-\delta_\star)}{\ln b} \Bigg\arrowvert_{g_\star^2, e_\star^2} = - \frac{c \epsilon}{N} (\delta - \delta_\star) + \mathcal{O} ((\delta - \delta_\star)^2),
\end{align}
with the constant $c>0$ as given below Eq.~\eqbetaFunctionDeltaExpanded\ in the main text. Importantly, the above flow equation now features a linear term~$\propto (\delta - \delta_\star)$, indicating the absence of a marginal parameter in the toy flow.
Consequently, the flow on the critical manifold $r=0$ is of usual power-law behavior towards the infrared stable fixed point.
This qualitatively change manifests itself also in all flowing exponents. They approach their true infrared values rapidly and no longer display intermediate drifting behavior.
This is illustrated in Fig.~\ref{fig:QuasiuniversalVsUniversal}, which shows the flowing fermion anomalous dimension $\eta_\psi$ as function of renormalization group scale $b^{-\epsilon}$. We reiterate that $\eta_\psi$ is related to the dynamical critical exponent $z$ via $\eta_\psi = 2 - z$, and as such governs the behavior of the specific heat $C \sim T^{d/z}$. In the toy model flow [Fig.~\ref{fig:QuasiuniversalVsUniversal}(a)], $\eta_\psi$ approaches zero already at $b^{-\epsilon} \simeq 10^{-3}$. By contrast, the Luttinger model flow [Fig.~\ref{fig:QuasiuniversalVsUniversal}(b)] features a wide quasiuniversal regime for $b^{-\epsilon} \lesssim 10^{-3}$ up to significantly below $b^{-\epsilon} = 10^{-6}$, characterized by a slow drift of $\eta_\psi = 2 - z$.
We have verified that all other flowing exponents display an analogous behavior.

%%%%%%%%%%%%%%%%%%%%%%%%%%%%%%%%%%%%%%%%%%%%%%%%%%%%%%%%%%%%%%%%%%%%%%%

%%%%%%%%%%%%%%%%%%%%%%%%%%%%%%%%%%%%%%%%%%%%%%%%%%%%%%%%%%%%%%%%%%%%%%%
% BIBLIOGRAPHY: FOR USE WITH BIBTEX
%%%%%%%%%%%%%%%%%%%%%%%%%%%%%%%%%%%%%%%%%%%%%%%%%%%%%%%%%%%%%%%%%%%%%%%
\bibliographystyle{longapsrev4-2}
\bibliography{QBT3D-Weyl}
%%%%%%%%%%%%%%%%%%%%%%%%%%%%%%%%%%%%%%%%%%%%%%%%%%%%%%%%%%%%%%%%%%%%%%%